# Electromagnon excitations in high temperature superconducting states

Yong-Jun Chen[1*]

[1]Department of Physics, Shaoxing University, Shaoxing, Zhejiang Province, 312000, China

*Corresponding author. Email: chenyongjun@usx.edu.cn

**Abstract**

In this article, we discover a fundamental excitation called electromagnon in the cuprate superconducting states. Doped holes render local inversion asymmetry which produces hidden Rashba spin-orbit coupling in copper-oxide plane. Rashba effect leads to Dzyaloshinskii-Moriya (DM) interaction between neighboring spins, which establishes a general spin Hamitonian combined with Heisenberg exchange interaction. The physical origin of spin fluctuations has been revealed from Rashba-induced DM term. Electromagnon excitations are found to result from spin fluctuations manifested as spin spiral states by hopping of doped holes in high temperature superconducting states. The frequency of electromagnon excitations, peak position of dielectric constant spectrum, decreases linearly with decrease of exchange interaction in various families of superconductors. Comparisons with superconductivity-induced $B_{1g}$ Raman peaks show good agreement and thus possibly provide the experimental evidences for electromagnon excitations in cuprates.

# I. Introduction

To unravel the mechanism for unconventional high temperature superconducting pairing is an outstanding challenge in condensed matter physics [1]. Isotopic effects indicate that the lattice vibration is involved in the pairing interactions of superconducting carriers [2, 3]. And experimental evidences [4, 5] showed that magnetism, that is, spins in superconducting plane (for example, spin fluctuations), should be included for the mechanism of the occurrence of superconductivity [1]. Taken together, the pairing mechanism should be related to both lattice motion and spin dynamics [2-5]. Thus, to discover the excitations related to both in high temperature superconductors will facilitate understanding the pairing mechanism for superconducting carries. Here, we note that electromagnon excitation, an electric-dipole active magnetic resonance, is such a possible candidate which is the hybrid excitation of phonon and magnon [6-16]. Electromagnon excitations are observed in multiferroics with magnetoelectric (ME) effect manifested by the magnetic field controlled ferroelectric polarization or electric field controlled magnetization [12, 13]. It appears as a peak in the imaginary part of dielectric constant spectrum $\varepsilon$ and is contributed by dynamic fluctuations of ferroelectric polarization $\vec{P} = e^*\vec{u}$, where $e^*$ and $\vec{u}$ are charge and lattice displacement respectively (Fig. 3(a)) [9-11, 13]. The electric field of electrons may excite electromagnon through magnetoelectric coupling [12, 13], similar to electron-phonon interaction in Bardeen-Cooper-Schrieffer theory [1]. Nevertheless, electromagnon is thought to exist in insulating multiferroics and observed in insulators, for example, perovskite $RMnO_3$ (R represents rare earth) [6-16].

Neutron scattering data revealed universal spin fluctuations in the copper oxide superconductors with hole doping [5, 17, 18]. The incommensurate wavevector $\delta$ of magnetic order evolve linearly with hole doping concentration $p = \delta$ in underdoped regime while $\delta$ keep constant in overdoped regime [5, 17]. The spin excitations demonstrate energy dispersion as shown in neutron scattering experiments [19-21]. The energy dispersion and the intensities in the spin fluctuations have been reproduced under the paradigm of spin spiral phase [22]. The spiral phase of spins has been shown for the ground state of spin structure in the cuprate superconductors due to the mobile vacancies [23]. Though the spin spiral phase has not been verified detailly due to the challenge of visualizing the individual arrangement of spins, split of one antiferromagnetic peak into four incommensurate peaks in inelastic neutron scattering provides compelling evidence for the spatial evolution of the spin spiral arrangements [17-21]. Same behavior of magnetic peak split also has been observed in other systems with spin spiral structure below Neel's temperature [24-26].

The electromagnon excitation is oscillation of electric polarization and occurs in noncollinear spiral magnets [6-16]. In the spiral structure of spins, spin current $\vec{j}_s \propto \vec{S}_i \times \vec{S}_j$ leads to electric polarization $\vec{P} \propto \vec{e}_{ij} \times (\vec{S}_i \times \vec{S}_j)$ (The unit vector $\vec{e}_{ij}$ connects adjacent spins $\vec{S}_i$ and $\vec{S}_j$) [27], which has been applied to spin spiral structure to explain the magnetoelectric (ME) effect [27]. For spin spiral states in copper oxide superconductors, ME effect also exists [28, 29] and electromagnon excitations can be predicted using mechanism of spin current [27]. The polarization oscillation of electromagnon excitations can be detected using Raman scattering as that have been shown for phonon in the superconducting states [30].

To demonstrate electromagnon excitations in cuprates, two difficulties are presented. Electromagnon excitations are related to the spin dynamics. Thus, a general spin Hamitonian should be established for doped cuprates while, heretofore, it has only been established in the parent insulators [31]. The spin Hamiltonian must explain the physical origin of spin fluctuations in cuprates. We solve these two problems by disclosing doping-induce hidden Rashba effect in CuO plane, which is our key contribution in this article.

## II. Dzyaloshinskii-Moriya interaction from Rashba effect

Doping creates holes in CuO plane. Doped holes prefer to reside primarily on oxygen sites and form Zhang-Rice singlet (ZRS) through hybridization with holes on Cu sites for hole-doped cuprate superconductors [32-34]. A hole on oxygen or a hole hopping from copper to oxygen induces the local inversion asymmetry (LIA) on copper (I in Fig. 1a). The LIA produces an electric field $E_x = 2.49 V Å^{-1}$ (calculated using point-like model) on the position of copper atom. The electric field from LIA will cause hidden Rashba effect of spin-orbit coupling on copper producing a spin torque on the spin in $d_{x^2-y^2}$ band on copper [35, 36]. Such hidden Rashba effect (R-2 effect in Ref. [35]) is similar to that from LIA of elements or crystal structure [36]. Rashba effect can be described using Rashba Hamitonian [35-36]:

$$H_R = \alpha_R (\hat{x} \times \vec{k}) \cdot \vec{\sigma} \qquad (1)$$

where $\alpha_R$ is Rashba constant, $\hat{x}$ is unit vector along CuO bond, $\vec{k}$ corresponds to momentum of electron and $\vec{\sigma}$ is Pauli matrix. The Rashba effect leads to spin splitting $\Delta\epsilon$ of $d_{x^2-y^2}$ band [36, 37]. Rashba constant $\alpha_R$ can be estimate using [37] $\alpha_R \sim \Delta\epsilon/k_f$, where $k_f$ is the wave vector on the Fermi surface. Rashba effect will leads to unique spin textures [35]. In the cell *i* with LIA, Rashba effect leads to a spin rotation of the electron at R*i* (Cu) that is polarized by Heisenberg interaction and the electron will propagate to another site R*j* (manifested as hole hopping) and interact with spins nearest to the site R*j*. Thus, Rashba effect of LIA will be transmitted to the whole CuO plane through the random mobility of holes. The interactions between two spins are mediated by mobile holes. This is similar to the magnetic interaction mediated by free electrons in the antiferromagnets [38]. Dzyaloshinskii-Moriya interaction (DMI) can be derived from the Rashba Hamitonian [38]. DMI from LIA is in the $z$ direction (in the inversion plane) normal to the CuO plane. According to the estimation in Rashba antiferromagnets [38], the micromagnetic form strength of DMI from Rashba effect can be approximately expressed as

$$D = \left(\frac{Ak_0^2}{4t}\right)\alpha_R \qquad (2)$$

where $A$ is a parameter characterizing the free carrier contribution to the exchange stiffness and $k_0 = \frac{2}{a}$ ($a$=3.8Å is lattice constant for Cu-O-Cu bond length in cuprates).

Presently, the measurement of Rashba constant in cuprates is not available. The energy of Rashba effect is part of the cohesive energy of the crystal [36]. Accurate calculation will show band splitting by Rashba effect. We estimate the Rashba constant of La$_{1.75}$Sr$_{0.25}$CuO$_4$ (p=0.25) from $d_{x^2-y^2}$ bond band splitting at Fermi Surface according to the accurate calculation of SCAN [39]. The band splitting $\Delta\varepsilon = |k|\alpha_R^{0.25} = 60.0 meV$ at nodal point $k = (0, 2/a)$ (M in nonmagnetic Brillouin zone [39]. Similar band spitting is seemly shown by ARPES for p=0.22 of La$_{1.78}$Sr$_{0.22}$CuO$_4$ in Fig. 6 of Ref. [40]) and we obtain $\alpha_R^{0.25}$=114.6meV·Å. From band structure of La$_{1.75}$Sr$_{0.25}$CuO$_4$, parameter $A = \frac{tJ_{sd}^2}{2\pi\varepsilon_F^2}(2 - \frac{\varepsilon_F^2}{16t^2})$ with $J_{sd} = 0.40 eV$ (a half of the gap between valence band and conduction band, $\varepsilon_F = 0.18 eV$ (from center of the gap), $t = 0.356 eV$ [38, 39, 41]. Substituting into Eq. 2, we obtain DMI strength $D$=12.32meV·Å$^{-1}$ of La$_{1.75}$Sr$_{0.25}$CuO$_4$ at p=0.25.

As shown in II of Fig. 1a, when two holes are located on the oxygen atoms on two sides of copper atom, LIA will disappear and the DMI will disappear on the center copper. When the concentration is relatively high, there is possibility that holes will be located nearest to each other on oxygen atoms along O-Cu-O bond. Thus the band splitting and Rashba constant are dependent on concentration of holes. Since the motion of holes is random along Cu-O-Cu bond, one hole can hop in four directions from Cu atom to oxygens. Hopping in *x* or *y* direction will contribute to the DMI distributed in *x* or *y* direction respectively. When two holes neighbor to each other as III in Fig. 1a, the electric field caused by the two holes will contribute to DMI distributed in *x* and *y* direction respectively.

Doped holes are primary located on oxygen sites [32-34]. The doped hole forms ZRS with the hole of d orbit on Cu [33]. The holes also hop by the form of ZRS [33]. As free carries, the two doped holes should not share one same Cu atom. To hop from one oxygen atom to another neighboring oxygen atom (in *x* or *y* direction) and form ZRS with the hole on another Cu d orbit, the neighboring Cu atoms should have unpaired $d_{x^2-y^2}$ holes for forming ZRS. Consequently, the doped holes should not be nearest or next nearest neighboring to each other (IV in Fig. 1a). For simple, we consider the mobile holes distributed on specific oxygen sites along Cu-O-Cu bond in *x* direction (or in *y* direction). Using Monto Carlo simulation, we calculate the possibility $P_0$ (one hole will not be nearest or next nearest neighbor to other holes shown by IV in Fig. 1a) for the holes on the specific oxygen atom in *x* direction along Cu-O-Cu bond at various doping levels. Shown in Fig. 1b is the doping-dependent $P_0$. The total Rashba energy in *x* direction $E = N\frac{p}{2}(P_0 \times 2\epsilon)$, where $N$ is the number of oxygen atom in O-Cu-O along *x* direction and $\epsilon$ is Rashba energy in one unit cell with LIA. Rashba energy $\epsilon = \alpha_R k_f$ for one cell with LIA. Doping dependent averaged Rashba constant can be expressed as $\alpha_R^p = {2pP_0\epsilon}/{k_f} = 2pP_0\alpha_R$. Using value of Rashba constant $\alpha_R^{0.25}$ at p=0.25 ($2pP_0 = 0.044$ in Fig. 1a), we can obtain $\alpha_R = 2604.5 meV \cdot Å$. The doping depedent $\alpha_R^p$ is shown in Fig. 1b. To evaluate doping dependent DMI strength, $J_{sd}$, $\varepsilon_F$ and $t$ from accurate band structure are needed. Since the band structure is vastly different from the different calculating methods [39] and systematic band structures are not available, these parameters varying with hole concentration are not clear.

Hereto we have unraveled that there exists DMI normal to the CuO plane between spins from hidden Rashba effect caused by hole doping. DMI prefers the spin spiral textures and will produce incommensurate spin fluctuations in hole-doped cuprates as shown by neutron scattering [5, 17]. On the contrary, in the electron-doped cuprates, the doped electrons reside on copper sites [34] and will not lead to LIA on the copper site in the unit cell. There is no hidden Rashba effect and thus spin fluctuations are commensurate [34]. Consequently, the hidden Rashba effect is compelling in the hole-doped cuprate superconductors.

For calculation of DMI, we also can estimate the DMI vector through the spatial distribution of spins by fitting DMI vector to spatial evolution of spin spiral texture (spin fluctuations) [42]. Inspired by neutron scattering results, similar to determination of DMI vector in multiferroics systerms [42, 43], the value of $D_{ij}$ is determined from the spin fluctuation parameter $\delta$ by $tan\theta = D_{ij}/J_1^{eff}$ [42, 43] to produce spiral state, where $D_{ij}$ is atomistic form of DMI, $J_1^{eff}$ is the first nearest neighbor exchange coupling (see Appendix) and $\theta = \delta\pi$. The conversion between atomistic form $D_{ij}$ and micromagnetic form $D$ of DMI is $D = \frac{r_{ij}D_{ij}}{\Delta r} = D_{ij}$ [44], where $r_{ij}$ and $\Delta r$ are distance between nearest spins and distance between two neighboring atoms within the chain, respectively. The DMI from spin spiral states of spin fluctuations is shown in Fig. 1b (D-spatial). DMI is proportional to the Rashba constant (Eq. 2). As shown in Fig. 1b, DMI shows similar evolution with Rashba constant against doping level. The parameter of exchange stiffness $A$ in Eq. 2 is doping-dependent. If we assume that $A$ is constant, the evolution of DMI is shown in Fig. 1b (D-r). The evolution trend agree reasonably well with each other though the value of predicted DMI is vastly different suggesting that *A* is variable with the doping level. Thus, we will adopt the method using spatial evolution of spin to estimate DMI because the accurate band structures of doped cuprates are not available presently due to the limited resolution.

### III. Spin fluctuations and spin spiral states

Doping of holes destroys the Neel's order of the parent insulator. The alignments of spins deviate the Neel's order and incommensurate with antiferromagnetic order. The orientations of spins in the CuO plane evolve with a wave vector $q = 2\delta$ as shown by neutron scattering experiments [23-26, 45]. The holes distribute randomly on the two-dimensional CuO plane [46]. The holes hop along Cu-O-Cu bonds. Spin spiral states have been predicted based on the *t-J* Hamitonian [23]. The vacancies of spins on Cu ions were assumed to create the "twist" on spin background [23] and such disordering deviates the spin configuration from antiferromagnetic order. The spin spiral develops. However, the physical essence of "twist" is not clear. As shown above, Rashba effect from the holes causes DMI between nearest spins which produces torque on spins, prefers a "twist" spin structure and thus explain the "twist" assumption of mobile holes [23]. Consequently, for hole doped cuprates, the spin Hamiltonian consists of a Heisenberg exchange term and a DMI term: $H = H_{ex} + H_{DMI}$, with

$$H_{ex} = J_1^{eff} \sum_{<i,j>} \vec{S}_i \cdot \vec{S}_j + J_2^{eff} \sum_{\ll i,l \gg} \vec{S}_i \cdot \vec{S}_l + J_3^{eff} \sum_{<<<i,i'>>>} \vec{S}_i \cdot \vec{S}_{i'} + J_4^{eff} \sum_{\lll i,l' \ggg} \vec{S}_i \cdot \vec{S}_{l'} \quad (3)$$

$$H_{DMI} = \sum_{<i,j>} \vec{d}_{ij} \cdot (\vec{S}_i \times \vec{S}_j) \quad (4)$$

We choose four effective exchange Hamiltonian $H_{ex}$, where $J_1^{eff}$, $J_2^{eff}$, $J_3^{eff}$, and $J_4^{eff}$ are spin-spin exchange interactions between first nearest neighbor （$<i,j>$）, second nearest neighbor （$\ll i,l \gg$）, third nearest neighbor （$<<<i,i'>>>$） and forth nearest neighbor （$\lll i,l' \ggg$） on Cu ions [47, 48], respectively. The determination of $J_\mu^{eff}$ ($\mu = 1,2,3,4$) is shown in the Appendix. Rashba-induced DMI is chosen to be $\vec{d}_{ij} = (0,0,d_z)$ to produce the spin spiral states in the CuO plane. The DMI components in CuO plane ($d_x$ and $d_y$) of $\vec{d}_{ij}$, weakly canting the spins out of the plane [49], are omitted. *Hereto, we have established a general spin Hamitonian to describe spin dynamics of doped cuprates.* The above model will produce incommensurate spin fluctuations or spin dynamics on CuO plane. Similar to determination of DMI vector in multiferroics systerms [42, 43], the value of $d_z$ is determined from the incommensurate wavevector $\delta$ of spin fluctuations by $tan\theta = d_z/J_1^{eff}$ [42, 43] to produce spiral state ($\theta = \delta\pi$). It agrees with that determined using Rashba constant as shown in Fig. 1(c).

The spin dynamics and electromagnon excitations are studied by numerically solving the Landau-Lifshitz-Gilbert (LLG) equation using the fourth-order Runge-Kutta method with projection onto a sphere surface [50, 51]. The LLG equation is given by [50]

$$\frac{\partial \vec{S}_i}{\partial \tau} = -\vec{S}_i \times H_i^{eff} + \frac{\alpha_G}{S}\vec{S}_i \times \frac{\partial \vec{S}_i}{\partial \tau} \qquad (5)$$

Where $\alpha_G = 0.04$ is the Gilbert-damping coefficient and $\tau$ is time. The exchange field $H_i^{eff}$ acting on the *i*th spin is derived from Hamiltonian $H$ using $H_i^{eff} = -\partial H/\partial \vec{S}_i$. The norm of the spin vector is $|\vec{S}_i| = 1/2$.

Equation (5) was discretized and implemented on the 50×50 mesh (m, n) (m stands for row and n stands for column) with periodical boundary conditions. The time step in the simulation is Δt=0.0001ps. We choose four families of superconductors: $La_{2-x}Sr_xCuO_4$ (LSCO, p=0.07, 0.15, 0.22, 0.30), $YBa_2Cu_3O_{6+y}$ (Y-123, p=0.10, 0.16), $HgBa_2CuO_{4+\delta}$ (Hg-1201, p=0.18) and $Bi_2Sr_2CaCu_2O_{8+\delta}$ (Bi-2212, p=0.0625, 0.082, 0.16, 0.165, 0.19, 0.21, 0.214, 0.246) at different doping levels.

We start from spiral configuration of spin at (m, n) grid [45]:

$$S_{mn} = (-1)^{m+n}\begin{pmatrix} S_{1x}\cos(m\delta 2\pi + \varphi_{1x}) + S_{2x}\cos(n\delta 2\pi + \varphi_{2x}) \\ S_{1y}\cos(m\delta 2\pi + \varphi_{1y}) + S_{2y}\cos(n\delta 2\pi + \varphi_{2y}) \end{pmatrix} \quad (6)$$

where $\varphi_{1x} = \varphi_{2x} + \Delta\varphi$, $\varphi_{2x} = 0.1$, $\varphi_{1y} = \varphi_{1x} - \pi$, $\varphi_{2y} = \varphi_{2x} - \pi$, $\Delta\varphi = \theta - 2\pi\delta(i-j)$, $\cos\theta = \frac{S_0^2 - S_{1x}^2 - S_{2x}^2}{2S_{1x}S_{2x}}$, $S_{1x} = S_0/2$, $S_{2x} = \sqrt{3}S_0/2$, $S_{1y} = S_{1x}$, $S_{2y} = S_{2x}$, $S_0 = 1/2$, $p = \delta \leq 0.15$. When $0.15 < p < 0.30$, the initial spin structure is same as that of $\delta = 0.15$. After enough relaxation using LLG equation (the relaxation also can start from the Neel's state), the spin spiral approaches to static spiral state (see Fig. 2 and supplemental video 1-4). *Thus, our model reproduces spin fluctuations in the cuprates and unravels the essence of the physical origin for the spin fluctuations in the cuprates.* This is one of the key contributions in this article.

### IV. Electromagnon excitations

To excite electromagnon, an electric pulse $\vec{E} \| z$ is applied. The time evolution of the spin structure and spontaneous polarization $\vec{P}_i$ due to spin current is traced by LLG equation [50]. The electromagnon spectrum $\text{Im}\varepsilon(\omega)$ is obtained from Fourier transformation of $\vec{P}_i(\tau)$ [50]. Also, the frequencies $\omega_{EM}$ of electromagnon are obtained by calculate the averaged period $T = 1/\omega_{EM}$. The spontaneous electric polarization due to the spin current can be calculated using

$$\vec{P}_i = \gamma_{ME} \vec{e}_{i,i+\hat{1}} \times (\vec{S}_i \times \vec{S}_{i+\hat{1}}) \qquad (8)$$

at bond site $i$ [50]. The magnetoelectric susceptibility $\gamma_{ME} = 4\pi\beta\gamma_M \langle\langle S \rangle\rangle_{av}$ [53], where $\gamma_M$ is magnetic susceptibility, $\beta$ is constant and $\langle\langle S \rangle\rangle_{av}$ is the averaged magnetization. In the superconducting regime, the magnetic susceptibility $\gamma_M \approx 0$ in the normal state and, in the superconducting transition, decreases to negative value ($\gamma_M < 0$) according to the susceptibility measurements [54]. Thus, we can expect $\gamma_{ME} \neq 0$ in the superconducting states from the proportional relation between $\gamma_{ME}$ and $\gamma_M$. Magnetoelectric coupling has been observed in the lightly underdoped $La_2CuO_{4+x}$ [55] while not found for the normal state in the superconducting regime [54] because of $\gamma_{ME} = 0$ due to $\gamma_M = 0$. We set magnetoelectric susceptibility $\gamma_{ME} = 10.0$ in the calculation (it not available in literatures). $\vec{e}_{i,i+\hat{1}}$ is the unit vector connecting the adjacent spins $\vec{S}_i$ and $\vec{S}_{i+\hat{1}}$ (Fig. 3(a)). Thus, the direction of polarization on the row is along $y$-axis (column) within $xy$-plane while on the column is along $x$-axis (raw) within $xy$-plane (Fig. 3(a)). To obtain the electromagnon excitations, the coupling between the polarization and the external exciting electric field $\vec{E}$ is $-\vec{E} \cdot \vec{P}_i = -\gamma_{ME}\vec{E} \cdot [\vec{e}_{i,i+\hat{1}} \times (\vec{S}_i \times \vec{S}_{i+\hat{1}})]$ [50], which regulates the DMI vector $\vec{d}_{ij}$ to $(\vec{d}_{ij} - \gamma_{ME}(\vec{E} \times \vec{e}_{i,i+\hat{1}}))$ in the $H_{DMI}$ and cants spins out of plane, where $\vec{E} = (E, 0, 0)$ according to the selection rule of spin current mechanism [14]. The delta-function mode of pulse is chose to be $-\gamma_{ME}(\vec{E} \times \vec{e}_{i,i+\hat{x}}) = (0, \frac{0.1 d_z \sin(10t-2)}{[(10t-2)\pi]}, 0)$ along $x$ direction and $-\gamma_{ME}(\vec{E} \times \vec{e}_{i,i+\hat{y}}) = (\frac{-0.1 d_z \sin(10t-2)}{[(10t-2)\pi]}, 0, 0)$ along $y$ direction ($\hat{x}$, and $\hat{y}$ represent adjacent point along raw ($x$) and column ($y$) directions, respectively). The parameters input into the model are summarized in Table 1 (see Appendix) for four families of superconductors at different doping levels.

Figure 3(b) and Fig. 4 shows typical electric polarization and electromagnon spectrums for different superconductors. We consider the polarization $P_n$ along $y$ direction on the row (the polarization along $x$ direction on the column has similar results of frequencies). The positions of the main peak $\omega_{EM}$, corresponding to the energy of electromagnon, depend on the doping level (Fig. 3(c)). The oscillation of polarization is almost anti-phase between two adjacent points and the sign of $\vec{P}_i$ on the neighboring points is different (Fig. 3(a)). The phase changes from point to point (Fig. 5 and Video 5). During the oscillation, the period fluctuates with time. The energy of electromagnon decreases with the increase of the doping level from around 15.0THz with p=0.0625 to 3.2THz with p=0.246 (Fig. 3c). As shown by the calculation of the electromagnon frequencies by point to point, the energy of electromagnon fluctuates with the coordinates of the grid points (Fig. 6(a)). We plot $\omega_{EM}$ verse $J_1^{eff}$ (Fig. 6(b)). $\omega_{EM}$ increases linearly with the increase of $J_1^{eff}$ manifesting the magnon characteristic of electromagnon. The nearest-neighboring exchange interaction $J_1^{eff}$ dominates the oscillation of electric polarization.

Electromagnon excitations can be measured through Raman scattering [15, 16]. The structure of CuO plane is tetrahedral with space group I4/mmm ($D_{4h}$). Electric wave function of $d_{x^2-y^2}$ orbit on $Cu^{2+}$ has the $D_{4h}$ point group symmetry. Distribution of charge $e^*$ around $O^{2-}$ is determined by the hybridization of $p_x$ or $p_y$ orbit of $O^{2-}$ and $d_{x^2-y^2}$ orbit of $Cu^{2+}$ (Fig. 3(a)). It is corresponding to the $B_{1g}$ irreducible representation. The oscillating displacement of O atom belongs to $B_{1g}$ mode which is Raman active and similar to the breathing mode of phonon, as shown in the unit cell (Fig. 3(a)). It corresponds to $B_{1g}$ mode in Raman spectrum, which will appear in the superconducting states. We survey the Raman measurements of cuprates in the superconducting states. Indeed, Raman peaks in the superconducting states have been revealed in many families of copper oxides superconductors [52, 56-64]. When the temperature drops across the transition temperature $T_c$ of superconductivity, a new strange continuum (superconducting peak) appears accompanying with inherent phonon peaks [52, 56-65]. We compare our above results of electromagnon with that of the $B_{1g}$ mode of superconductivity-induced Raman peaks. The frequencies $\omega_{EM}$ are normalized using $kT_c$ ($k$ is Botlzman constant) and Fig. 7 plots the doping evolution of normalized frequencies. The comparison shows good agreement between our electromagnon and $B_{1g}$ mode of superconductivity-induced Raman peaks (Fig. 7). In addition, electromagnon peaks and superconductivity-induced peaks in Raman spectral demonstrates similar evolution trend with pressure [15, 60]. Thus, it is possible that the superconductivity-induced $B_{1g}$ peaks origin from the Raman scattering of electromagnon excitations and provide experimental support for our calculations. This suggests that possibly electromagnon is related to the superconductivity in cuprates since the superconductivity-induced peaks has been attributed to the superconducting gap [52, 59, 61-64]. Whether electromagnon excitations enter the mediation of cooper pairing is the future research. Changing the polarization of light, we will obtain different modes of electromagnon excitations.

## V. Conclusion

In summary, we have demonstrated the occurrence of the electromagnon excitations in high temperature superconducting states by introducing Rashba spin-orbit coupling and spin fluctuations has been explained by a general spin Hamitonian combined with a DMI term. Superconductivity-induced $B_{1g}$ peaks in Raman measurements of superconducting cuprates provide the experimental evidence for the electromagnon in superconducting cuprates. Electromagnon excitation will be a promising candidate for pairing interaction due to the coupling interaction with electrons. We expect that electromagnon excitations also will be observed universally in other superconductors, for example, iron-based superconductors and heavy-fermions superconductors with spin fluctuations.

## Acknowledgment

This work was supported by ZJNSF (No. LY16A040003) and NSFC (No. 11204181).

# APPENDIX

## Calculation of $J_\mu^{eff}$ ($\mu = 1,2,3,4$) in $H_{ex}$

The effective exchange parameters $J_1^{eff}$, $J_2^{eff}$, $J_3^{eff}$, and $J_4^{eff}$ are determined following Ref. [47]: $J_1^{eff} = \left(\frac{4t^2}{U}\right)\left(1 - \frac{6t^2}{U^2}\right) - 2\left(\frac{80t^4}{U^3}\right)S^2$, $J_2^{eff} = \frac{4(t')^2}{U} - 16\frac{t^4}{U^3}$ (S=1/2), $J_3^{eff} = \frac{4t^4}{U^3} + \frac{4(t'')^2}{U}$, $J_4^{eff} = \frac{4t'^2 t^2}{U^3}$, where U is the effective Hubbard U, $t$, $t'$ and $t''$ are parameters in $t - t' - t'' - U$ model [48]. The parameters $t$, $t'$ and $t''$ are determined by fitting the Fermi surface using two-dimensional single-band tight-binding model [66]:

$$\varepsilon_k = \varepsilon_0 - 2t(cosk_x a + cosk_y a) - 4t' cosk_x a cosk_y a -$$
$$2t''(cos2k_x a + cos2k_y a) \qquad (7)$$

where $k_x$ and $k_y$ are coordinates of Fermi Surface in momentum space, and, $\varepsilon_k$, $\varepsilon_0$, and $a$ are Fermi energy, constant and crystal constant. We have $t'' = -t'/2$ for all doping levels. The decrease of Heisenberg exchange interaction $J_s$ with increase of doping level has been shown experimentally in Ref. [67-70] and indicated clearly by the doping dependence of paramagnon dispersion [71] and two-magnon peak energy $E_{max} \approx 3J_s$ [57, 72]. The effective U is determined by doping-dependent Heisenberg exchange interaction $J_s$ by $J_s(p) = \frac{4t^2}{U}$ [73], and $J_s(p)/J_s(0)$ is adopt from Ref. [73] assuming zero electronic anisotropy of material (purely two dimensional case, $\Gamma/W = 0$, V=0.50 in Ref. [73], see Fig. 8).

To calculate $J_\alpha^{eff}$ ($\alpha = 1,2,3,4$) in $H_{ex}$, we choose, for $p = 0$, $J_s(0) = 126 meV$, $U = 2.8 eV$, $t = 0.30 eV$ for Y-123 [74, 75]; $J_s(0) = 147 meV$, $U = 3.33 eV$, $t = 0.35 eV$ for Bi-2212 [76, 77]; $J_s(0) = 145 meV$, $U = 3.42 eV$, $t = 0.35 eV$ for Hg-1201 [78, 79]; and $J_s(0) = 136 meV$, $U = 3.65 eV$, $t = 0.35 eV$ for LSCO [78, 79]. The values of $t$ are fixed and $U's$ vary with doping. The data of Fermi surface of various families of superconductors can be found in Ref. [79-81] for Y-123, Ref. [82-84] for Bi-2212, Ref. [85] for Hg-1201, and Ref. [66] for LSCO and then is fitted using Eq. 7 with coordinates ($k_x$, $k_y$) of three or more points on the Fermi surface. The parameters are summarized in Table 1.

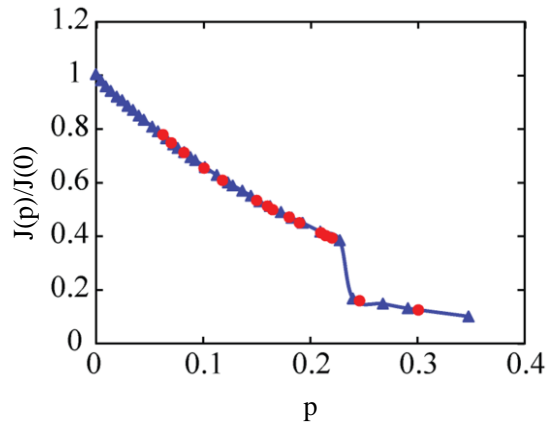

Figure 8 Doping-dependent effective exchange interaction of cuprates. Circle(●): values used in this work. Triangle (▲): values adopted from Ref. [73].

Table 1 Parameters input for numerical solvation of Equation (1), (2), and (4). All of the $J_4^{eff}$'s are negligible and have been set to zero in the calculation. $T_c = T_{c,max}[1 - 82.6(p - 0.16)^2]$ [52], where $T_{c,max}$ is the maximum $T_c$.

| Name | p | J(p) (meV) | U (eV) | t (eV) | t' (eV) | t'' (eV) | $J_1^{eff}$ (meV) | $J_2^{eff}$ (meV) | $J_3^{eff}$ (meV) | $T_c$ (K) |
|---|---|---|---|---|---|---|---|---|---|---|
| **Y-123** | 0 | 126.00 [73] | 2.8 [78] | 0.30 [73] | — | — | — | — | — | — |
| **Y-123** | 0.1 | 82.35 | 4.37 | 0.30 | -0.0810 | 0.0405 | 76.15 | 4.45 | 1.89 | 60[79] |
| **Y-123** | 0.16 | 64.47 | 5.58 | 0.30 | -0.0930 | 0.0465 | 61.49 | 5.45 | 1.73 | 93[80] |
| **Bi2212** | 0 | 147.00 [75] | 3.33 | 0.35 [76] | — | — | — | — | — | — |
| **Bi2212** | 0.0625 | 114.41 | 4.28 | 0.35 | -0.0606 | 0.0303 | 102.18 | 0.37 | 1.62 | 20.19 |
| **Bi2212** | 0.082 | 104.60 | 4.68 | 0.35 | -0.0578 | 0.0289 | 95.26 | 0.51 | 1.30 | 46.76 |
| **Bi2212** | 0.118 | 89.43 | 5.48 | 0.35 | -0.0840 | 0.0420 | 83.59 | 3.69 | 1.65 | 80.3 |
| **Bi2212** | 0.16 | 75.22 | 6.51 | 0.35 | -0.0574 | 0.0287 | 71.75 | 1.15 | 0.72 | 94[83] |
| **Bi2212** | 0.165 | 73.32 | 6.68 | 0.35 | -0.0945 | 0.0473 | 70.10 | 4.54 | 1.54 | 93.8 |
| **Bi2212** | 0.19 | 66.06 | 7.42 | 0.35 | -0.0539 | 0.0270 | 63.71 | 0.98 | 0.54 | 87.01 |
| **Bi2212** | 0.21 | 60.68 | 8.08 | 0.35 | -0.0525 | 0.0263 | 58.86 | 0.91 | 0.46 | 78[83] |
| **Bi2212** | 0.214 | 59.09 | 8.29 | 0.35 | -0.0900 | 0.0450 | 57.41 | 3.48 | 1.08 | 71.36 |
| **Bi2212** | 0.246 | 23.39 | 20.95 | 0.35 | -0.0935 | 0.0467 | 23.29 | 1.64 | 0.42 | 36.57 |
| **Hg1201** | 0 | 145.00 [77] | 3.38 [78] | 0.35 | — | — | — | — | — | — |
| **Hg1201** | 0.18 | 68.27 | 7.18 | 0.35 | -0.0511 | 0.0256 | 65.67 | 0.81 | 0.53 | 92[84] |
| **LSCO** | 0 | 136.00 [76] | 3.65 [77] | 0.35 | — | — | — | — | — | — |
| **LSCO** | 0.07 | 101.76 | 4.82 | 0.35 | -0.0595 | 0.0298 | 93.16 | 0.79 | 1.27 | 13.9 |
| **LSCO** | 0.15 | 72.52 | 6.76 | 0.35 | -0.0525 | 0.0263 | 69.41 | 0.85 | 0.60 | 41 [66] |
| **LSCO** | 0.22 | 53.50 | 9.16 | 0.35 | -0.0455 | 0.0228 | 52.25 | 0.59 | 0.30 | 29.51 |
| **LSCO** | 0.3 | 16.96 | 28.89 | 0.35 | -0.0420 | 0.0210 | 16.92 | 0.23 | 0 | — |

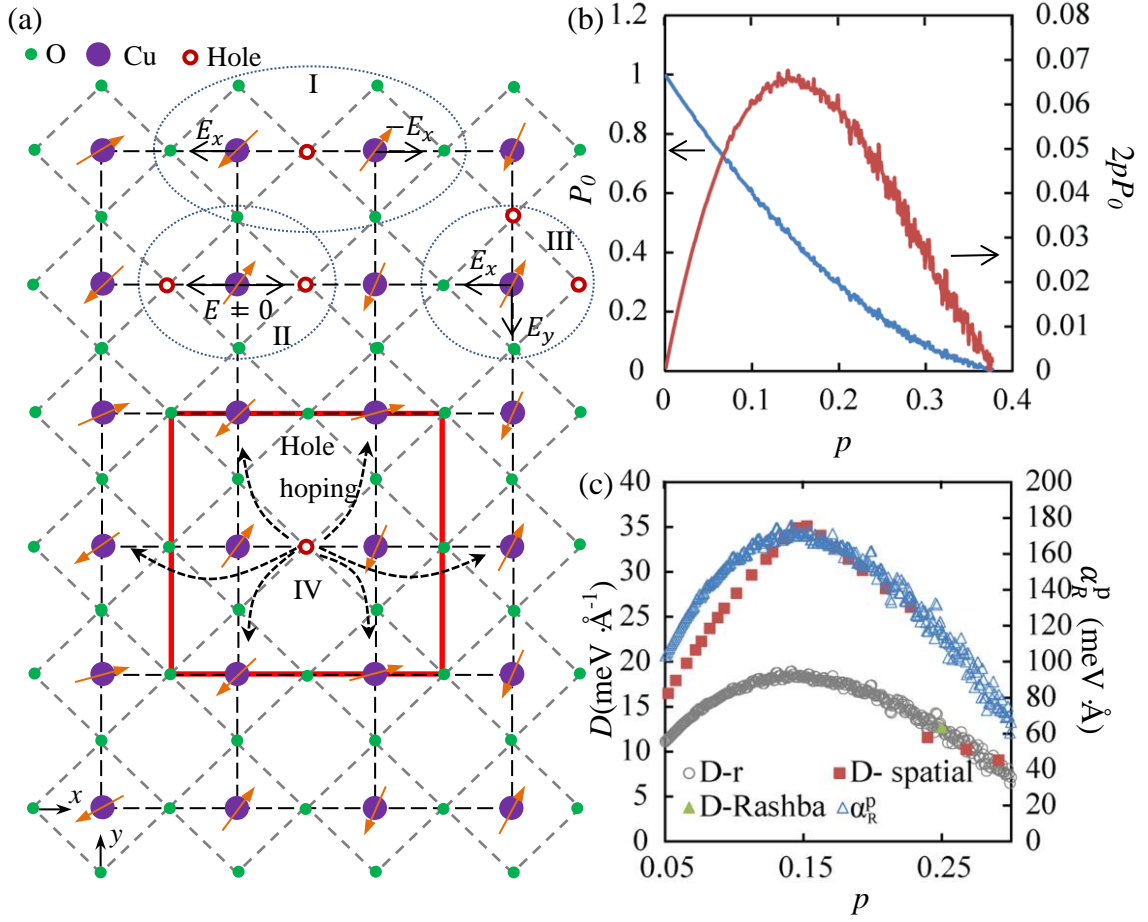

Figure 1 Rashba effect of a mobile hole and DMI. (a) Schematic of hole causing local spatial asymmetry. In the case of IV, the red square represents unit cell of oxygen in $x$ direction. To form ZRS and facilitate hole hoping, the doped holes should not reside on the sides of the square. (b) Monte carlo simulation of doping-dependent probability for a hole which is not nearest-neighbor to other holes (IV in (a)) on oxygen atoms and its product with doping level p. (c) Doping-dependent DMI strength and averaged Rashba constant $\alpha_R^p$.

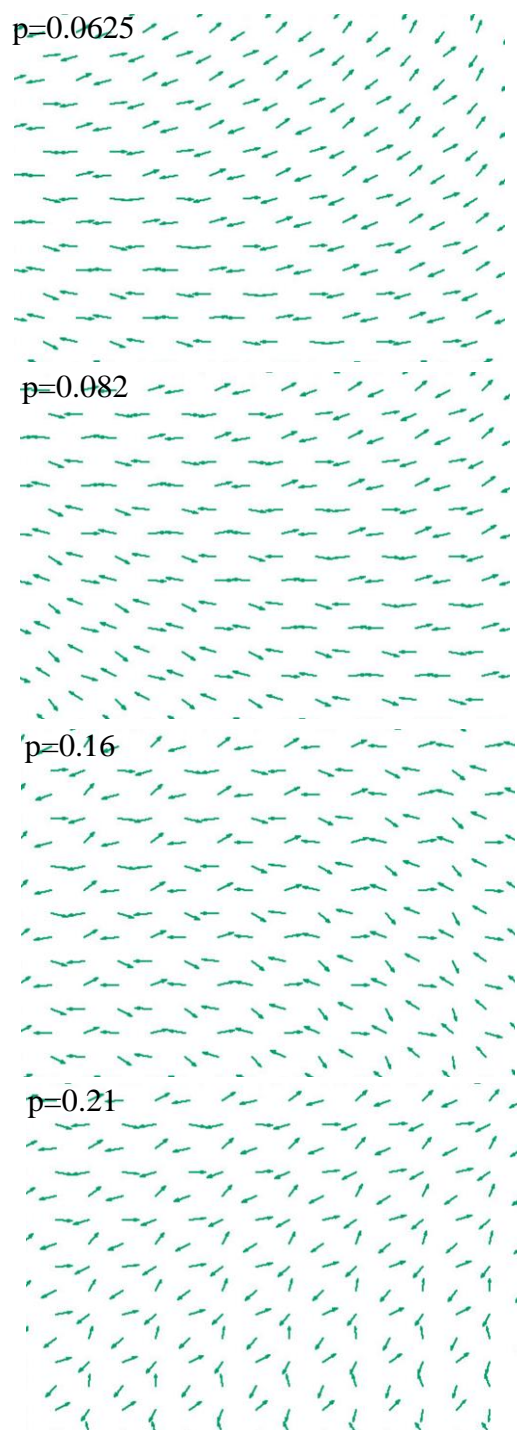

Figure 2 Spiral structure of spins on CuO plane. The directions of spins do not change and keep the static antiferromagnetic spiral state. The arrows represent the directions of spins. The starting point is on the grid position of Cu ions. For details, see videos 1-4 in the supplemental information.

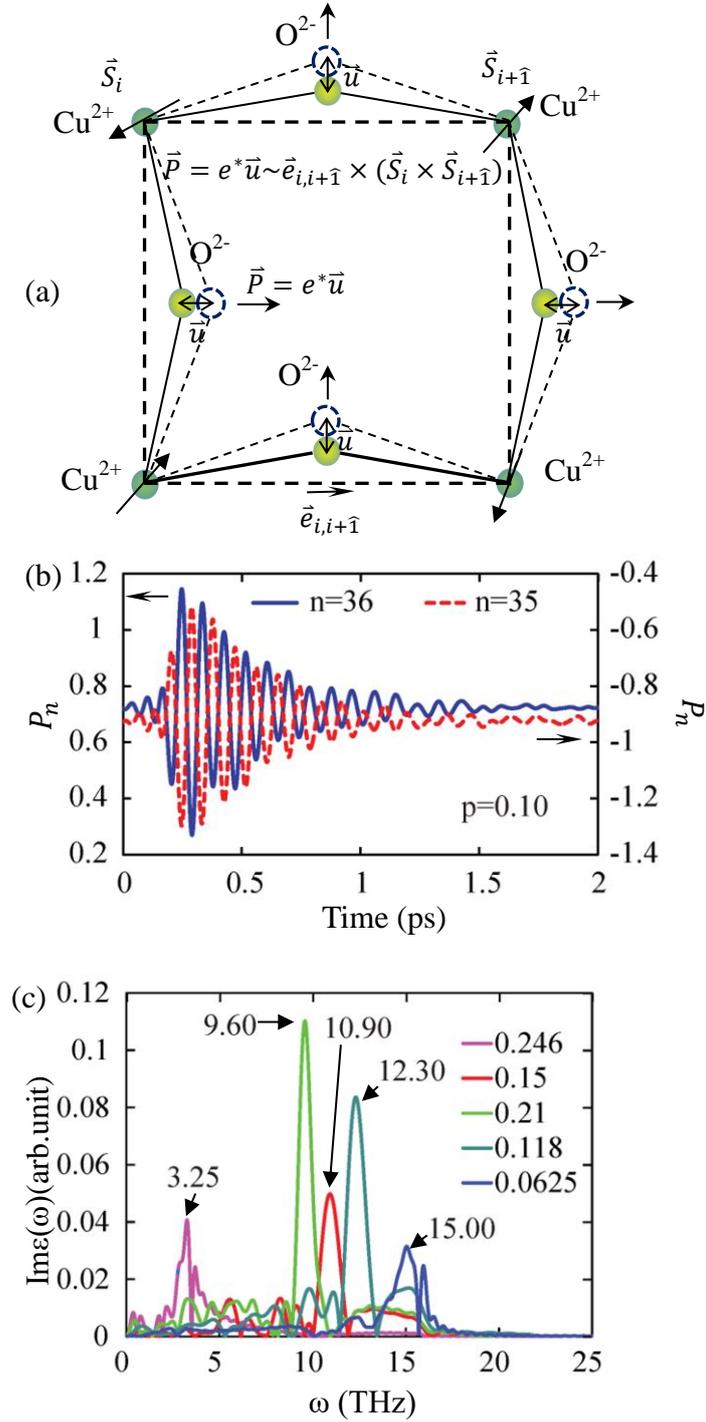

Figure 3 Electromagnon excitations. (a) schematic of unit cell in CuO plane and motion of atom O. (b) Oscillation of polarization on the Cu-O bond (m=14). (c) Spectral of Imε(ω) from Fourier transformation. Im(ε) is the dielectric response. The doping levels and the frequencies are indicated in the figure.

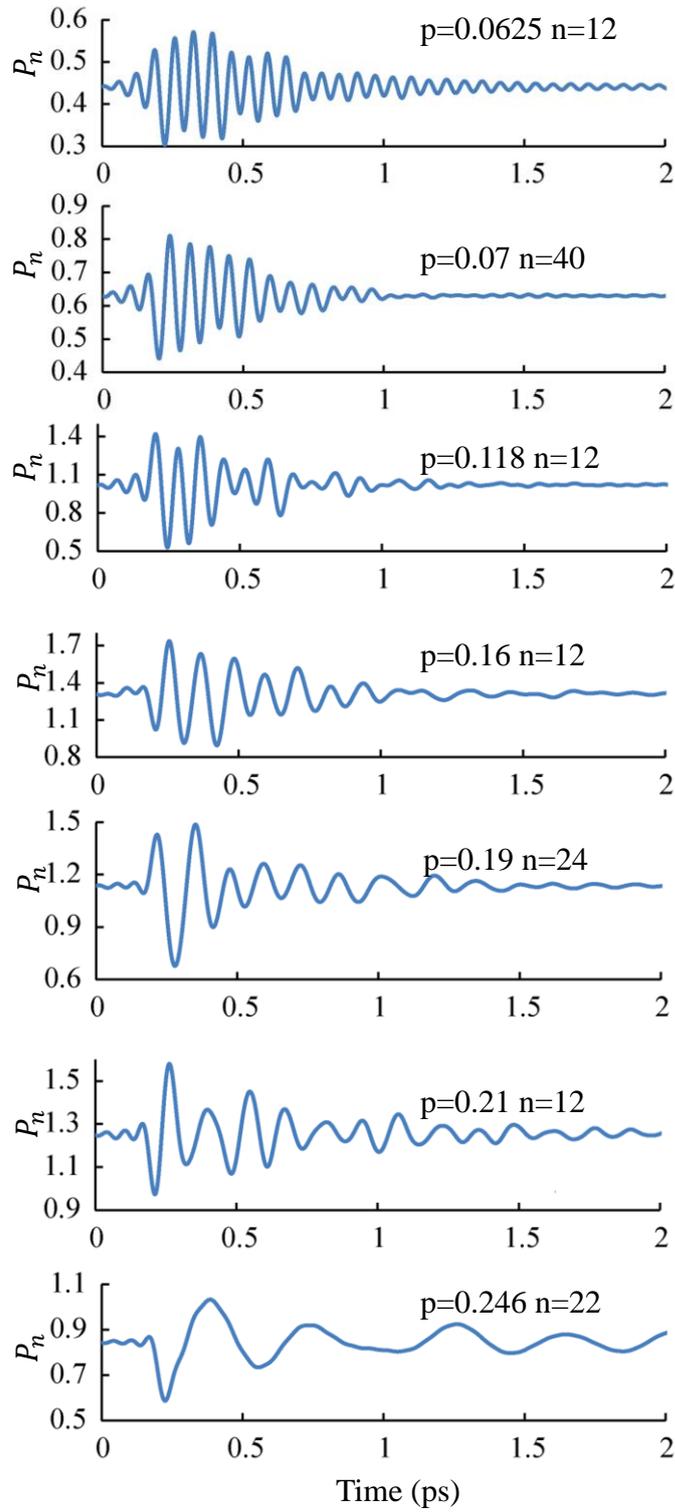

Figure 4 Typical oscillations of polarization at various doping levels (m=14).

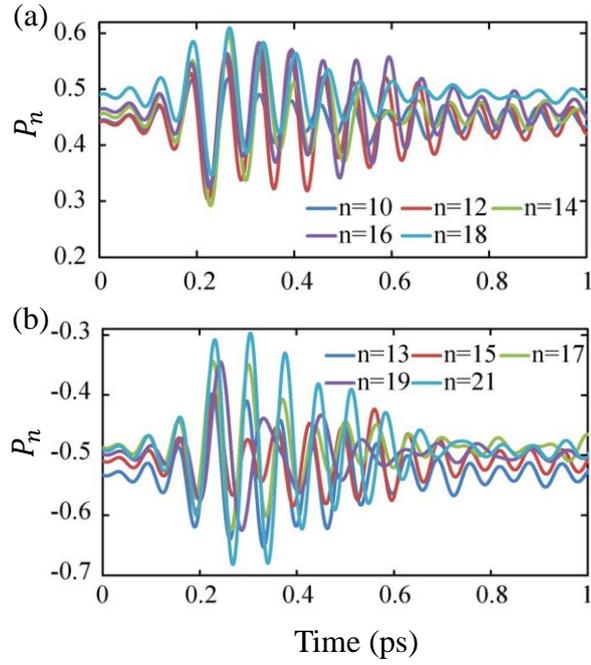

Figure 5 Oscillations of polarization at points with even and odd numbers n (m=14). (a) Even number n. (b) Odd number n. The doping level is p=0.16. For more details, see Video 5.

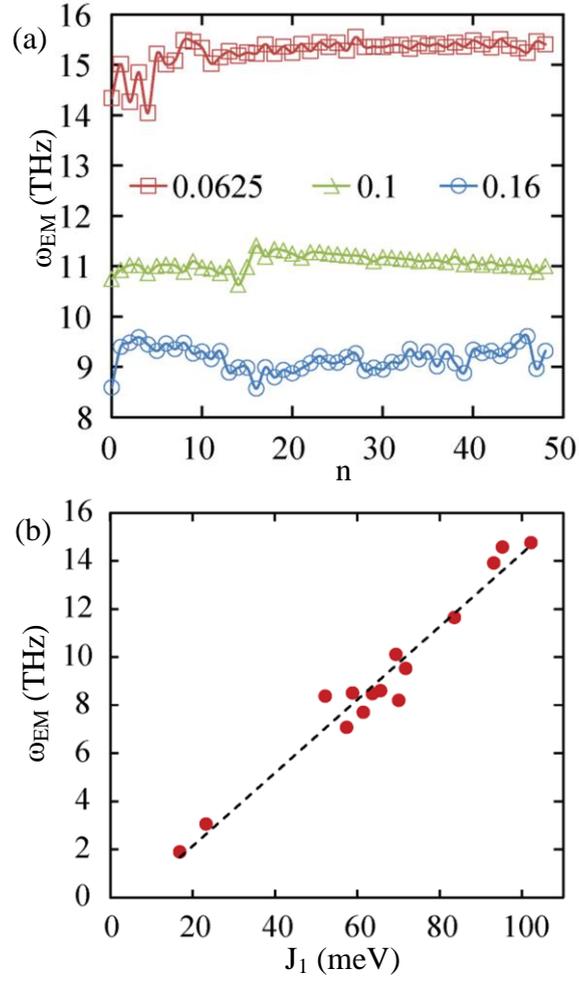

Figure 6 Doping-dependent frequencies of electromagnon excitations. (a) Doping-dependent frequencies vary with grid number n (m=14). The doping levels are indicated in the figure. (b) Linear relationship between frequencies and exchange interaction $J_1$.

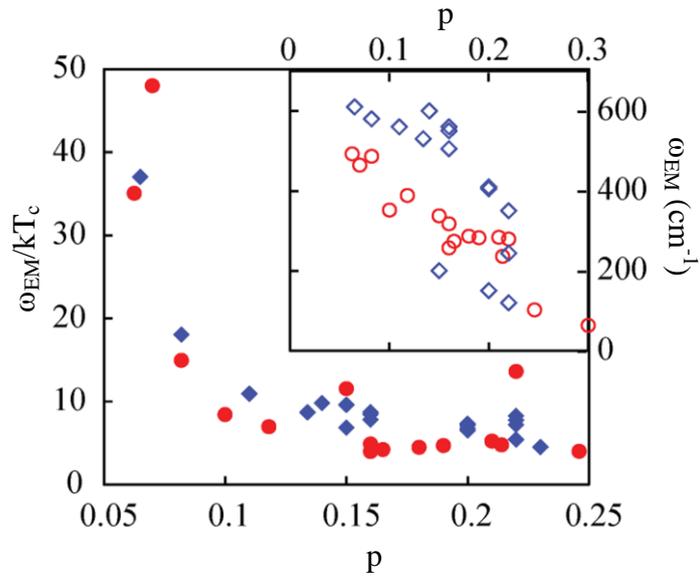

Figure 7 Comparisons between electromagnon and superconductivity-induced $B_{1g}$ peaks of cuprates. The frequencies are normalized using $kT_c$. In the inset is the comparison of the frequencies varying with doping level. Circle (●○) represents this work and diamond (♦◊) is superconductivity-induced $B_{1g}$ peaks for Bi-2212 [56], $Bi_2Sr_2(Y_{1-x}Ca_x)Cu_2O_{8+\delta}$ [52], $Bi_2Sr_2Ca_{1-x}Y_xCu_2O_{8+\delta}$ [57], Y-123 [57, 59], LSCO [57, 59].